\documentclass{aa}
\usepackage{epsfig}
\usepackage{graphicx}
\usepackage{longtable}
\begin{document}

\title{\ion{Mg}{ii} chromospheric radiative loss rates\\ 
in cool active and quiet stars}

\author{D. Cardini}

\offprints{Daniela Cardini,
\email{cardini@rm.iasf.cnr.it} }

\institute {Istituto di Astrofisica Spaziale e Fisica Cosmica, CNR,
Via del Fosso del Cavaliere 100, 00133 Roma, Italy}

\date{Received / Accepted}

\authorrunning{D. Cardini}

\titlerunning{\ion{Mg}{ii} emission fluxes in late type stars.}

\abstract{ 
The \ion{Mg}{ii} {\it k} emission line is a good indicator of the level of
chromospheric activity in late--type stars.
We investigate the dependence of this activity indicator on fundamental
stellar parameters.  
To this purpose we use {\it IUE} observations of the
\ion{Mg}{ii} {\it k} line in 225 late--type stars of luminosity classes I--V,
with different levels of chromospheric activity.
We first re--analyse the relation between \ion{Mg}{ii} {\it k} line luminosity 
and stellar absolute magnitude,
performing linear fits to the points. 
The ratio of \ion{Mg}{ii} surface flux to total surface flux is
found to be independent of stellar luminosity for evolved stars
and to increase with decreasing luminosity for dwarfs.
We also analyse the \ion{Mg}{ii} {\it k} line surface flux -- metallicity
connection. The \ion{Mg}{ii} {\it k} emission level turns out to be
not dependent on metallicity.
Finally, the \ion{Mg}{ii} {\it k} line surface flux -- temperature relation
is investigated by treating separately, for the first time, a large sample of
very active and normal stars. 
The stellar surface fluxes in the {\it k} line of normal stars are
found to be strongly dependent on the temperature and slightly dependent 
on the gravity,
thus confirming the validity of recently proposed models.
In contrast, data relative to RS CVn binaries and BY Dra stars, 
which show very 
strong chromospheric activity, are not justified in the framework
of a description based only on acoustic waves and uniformly distributed
magnetic flux tubes so that they require more detailed models.

 }
\maketitle

\section{Introduction}
\label{sec:intro}

The \ion{Mg}{ii} {\it h} and {\it k} resonance lines as well as 
the \ion{Ca}{ii} {\it H} and {\it K} lines have played and continue to play
an important role in the understanding of physical conditions
in the chromospheres of late-type stars.
In particular the \ion{Mg}{ii} emission lines, which are formed at 
higher temperatures and greater heights than the \ion{Ca}{ii} lines, 
are important tools for the diagnosis of chromospheric properties and,
thanks to the large abundance 
of Magnesium, are good indicators of the total chromospheric
radiative loss (\cite{linskyayres}).
In addition, the photospheric background around the region of
2800 \AA\ being small or negligible compared to the chromospheric flux,
the problem of photospheric background subtraction is considerably
reduced with respect to \ion{Ca}{ii}. 

For all these reasons, \ion{Mg}{ii} emission lines have been widely
studied both in the Sun and in late-type stars (\cite{basri}; 
\cite{cerruti}; (\cite{mathiou}); \cite{elgaroy}; \cite{buchholz}).

The presence of resonance emission lines in late type stars
is generally attributed to chromospheric temperature 
gradients generated
by nonradiative heating (\cite{ayres}). Two different types
of mechanisms have been identified as responsible for chromospheric heating:
hydrodinamic (acoustic and pulsational waves) and magnetic mechanisms 
(\cite{narain}; \cite{buchholz}; \cite{ulmsch}; \cite{fawzya}).

The heating by acoustic waves appears to be the basic mechanism 
generating a {\it basal} flux 
strongly dependent on the colour of the star, but independent 
of luminosity class (\cite{oranie}; \cite{schrijver}; \cite{rutten}).

The fact that stars with the same fundamental stellar parameters, like
effective temperature, luminosity, and metallicity show
a broad range of chromospheric activity is generally attributed to different
fractions of the stellar surface covered by magnetic fields (\cite{rutten};
\cite{baliun1}). 
Indeed, high-resolution observations of the Sun show considerable spatial 
variation in \ion{Mg}{ii} fluxes which become larger
in regions of higher magnetic activity.

Several semiempirical and theoretical models of stellar
chromospheres have been proposed (see \cite{ayres}; 
\cite{ulmsch}; \cite{fawzya}).
Theoretically predicted {\it basal} emission fluxes are in good
agreement with minimal observed emission fluxes. 
On the other hand, the modeling of magnetic 
surface structure, which is responsible for the 
{\it excess} flux, roughly accounts
for the observed line emission fluxes in \ion{Ca}{ii}, but does not explain
the large emission fluxes in \ion{Mg}{ii} (\cite{buchholz}; \cite{fawzyb}).

To test our understanding of the generation,
propagation and dissipation of both mechanical and magnetical
energy fluxes, 
it is useful to provide an empirical estimate of the amount of chromospheric 
heating in a large sample of stars covering a wide range of stellar 
properties. 

In this paper we present the results of the 
analysis of the \ion{Mg}{ii} {\it k} $\lambda$ 2796.34 \AA\ emission line 
of 225 stars, with spectral type F6--M6 and luminosity class I--V,
as measured from a large number
(more than 1400) of {\it IUE} spectra.
Our sample contains stars with different level of activity, namely 
dwarfs, very active main sequence stars like BY Dra,
giants and supergiants, and RS CVn, chromospherically active
evolved stars.

The interest of our analysis mainly lies in the following points:
\begin {enumerate}
\item The method used to measure intensities of lines is the same for 
a very large sample of different types of stars.
\item The absolute flux scale is the same for all our stars and is 
accurate due to the use of high precision measurements of distances.
\item The effective temperatures and bolometric corrections 
have been determined fron $B-V$ colours using unique accurate 
numerical relations.
\end {enumerate}

The paper is organized as follows.
In Sec. \ref{sec:sample} and \ref{sec:proces}
we describe our sample of stars, the {\it IUE} observations 
and their reduction to absolute surface flux units. 
In Sec. \ref{sec:discu} we search for trends of estimated 
chromospheric radiative loss rates 
with basic parameters of stars such as effective temperature $T_{eff}$,
luminosity and metallicity. 
In Sec. \ref{sec:interp} we compare our 
observations to avalaible theoretical models and our conclusions are
given in Sec. \ref{sec:concl}.

\section{Sample of stars}
\label{sec:sample}

The data base used in the present paper was generated starting from 
the original sample of 230 normal stars used in a study of the Wilson--Bappu 
effect by \cite{cassat} (hereafter Paper I).
It was a sample of stars with good quality \ion{Mg}{ii} observations 
from {\it IUE} (see Section \ref {sec:proces}) 
and known parallaxes from {\it HIPPARCOS} with luminosity classes ranging
from I to V. 
Since for the purpose of the present paper we need accurate
determination of spectral type, colour and bolometric correction, 
we have further selected the original sample 
in order to avoid 
indetermined spectral classification, large
errors on $B-V$ ($>$ 0.1) and too large bolometric corrections ($<$ -2.3) 
(see Section \ref {sec:proces}) because they are too uncertain.
We obtained a final sample of 173 nominally quiet stars.

In addition to normal stars, we consider here a sample of 29 
very active evolved stars of the RS CVn type. They have been 
selected 
following the criteria given in \cite{cardini} (hereafter Paper II) 
and further selected following the criteria illustrated above. 

Finally, we added a sample of BY Dra stars wich consist of both
single and binary dwarf stars having strong
\ion{Ca}{ii} H and K emission lines. We have included
all objects appearing as BY Dra stars in the SIMBAD data base for which
{\it IUE} high resolution long-wavelength spectra and {\it HIPPARCOS}
parallaxes were available. As selection criteria, we have adopted the same
as Paper I and Paper II and described above. 
These constitute our sample of 23 chromospherically active main 
sequence stars with spectral types F to K. 

The resulting sample (225 stars) consists of an accurately selected, 
uniformly reduced and homogeneously calibrated data set.

\section{Observations and data processing}
\label{sec:proces}

\begin{table*}
\caption{\ion{Mg}{ii} {\it k} line fluxes for BY Dra stars. For each object
we give the Hipparcos number, the star name,
the spectral type and luminosity class, 
the reference for spectral type, the $B-V$ color index, 
the absolute visual magnitude, the log of the surface \ion{Mg}{ii} 
{\it k} flux, the metallicity, and the indication of binarity.
For objects with double--lined spectra, the last column indicates which one of the
stellar components is the active star.}
\begin{flushleft}
\begin{tabular}{ r l l c c c c c c c }
\hline \hline \noalign{\smallskip}

HIP&Name&$Sp.Type$ & ref& $B-V$& $M_{\it V}$&  $log~F_{\ion{Mg} {ii}}$&[Fe/H] &bin& $Active~Star$\\
\noalign{\smallskip}
\hline \noalign{\smallskip}
  1803 & BE Cet       & G4 V     &1&  0.66 &   4.84 &   6.15 & -0.01 & &\\
 11964 & CC Eri       & K7 V/M3 V&2&  1.39 &   8.58 &   6.35 &  ---  &*&\\
 16537 & $\varepsilon$ Eri & K2 V&3&  0.88 &   6.18 &   6.04 & -0.09 & &\\
 19855 & V891 Tau     & G6 V     &3&  0.68 &   5.34 &   6.28 &  ---  & &\\
 19934 & V984 Tau     & K0 V     &4&  0.81 &   5.59 &   6.12 &  ---  & &\\
 20237 & V986 Tau     & G0 V     &4&  0.56 &   4.20 &   6.30 &  ---  & &\\
 20485 & V989 Tau     & K5 V     &4&  1.23 &   7.09 &   5.86 &  ---  & &\\
 20553 & V895 Tau     & G1 V     &4&  0.60 &   4.32 &   6.36 &  ---  & &\\
 20719 & V906 Tau     & G1 V     &4&  0.65 &   4.73 &   6.20 &  ---  & &\\
 20815 & V993 Tau     & F8 V     &4&  0.54 &   4.11 &   6.16 &  ---  & &\\
 20890 & V918 Tau     & G8 V     &4&  0.74 &   5.14 &   6.38 &  ---  & &\\
 20899 & V920 Tau     & G2 V     &4&  0.61 &   4.45 &   6.23 &  ---  & &\\
 21482 & V833 Tau     & K3 IV    &3&  1.10 &   6.84 &   6.40 &   --- &*&\\
 30630 & OU Gem       & K3 V/K5 V&5&  0.94 &   5.95 &   6.32 &   --- &*&Both\\
 46816 & LQ Hya       & K2 V     &2&  0.93 &   6.50 &   6.31 &   --- & &\\  *
 46843 & DX Leo       & G9 V     &3&  0.78 &   5.80 &   6.28 &  ---  & &\\  *
 57269 & V838 Cen     & K1 V/K2 V&2&  0.91 &   5.47 &   6.33 &  ---  &*&Both\\
 71631 & EK Dra       & G5 V     &3&  0.63 &   4.95 &   6.53 &  ---  & &\\  *
 72659 & 37 Boo       & G7 V     &3&  0.72 &   5.41 &   6.22 & -0.15 & &\\  
 81300 & V2133 Oph    & K0 V     &3&  0.83 &   5.82 &   5.90 &  0.01 & &\\
 82588 & V2292 Oph    & G8 V     &3&  0.75 &   5.51 &   6.25 &  ---  & &\\
 83601 & V2213 Oph    & F9 V     &3&  0.58 &   4.45 &   6.07 & -0.19 & &\\
 91009 & BY Dra       & K4 V/K7.5 V&5&1.27 &   7.12 &   6.25 &  ---  &*&Both\\
\noalign{\smallskip} \hline \noalign{\smallskip}
\end{tabular}
\end{flushleft}
\par
\vspace{0.3cm} \noindent {\it Notes}:{
(1) \cite {cutispotoa};
(2) \cite {cutispotob};
(3) \cite {gray};
(4) Strasbourgh Data Center (SIMBAD);
(5) Catalog of Active Binary Stars (CABS) by \cite {strass}.\\
The last two Columns are according to CABS except for V838 Cen 
(see Cutispoto 1998)}.
\label{tab:data1}
\end{table*}
\begin{table*}
\caption{\ion{Mg}{ii} {\it k} line fluxes for RS CVn stars. 
For each object we give the Hipparcos number, the star name,
the spectral type and luminosity class, the orbital period, 
the $B-V$ color index, the absolute visual magnitude, 
the log of the surface \ion{Mg}{ii} {\it k} flux, and the metallicity.
For objects with double--lined spectra, the last column indicates which one of the
stellar components is the active star.}
\begin{flushleft}
\begin{tabular}{ r l l c c c c c c }
\hline \hline \noalign{\smallskip}

HIP&Name&$Sp.Type$ & $P_{orb}$(days)& $B-V$& $M_{\it V}$&  $log~F_{\ion{Mg} {ii}}$&[Fe/H] &$Active~Star$\\
\hline \noalign{\smallskip}
  4157 & CF Tuc       & G0 V/K4 IV      &   2.8  & 0.68 &    2.93 &   6.82&---  & Cool    \\
  9630 & XX Tri       & K0 III          &  24.0  & 1.11 &    1.92 &   6.74&---  &         \\
 10280 & TZ Tri       & F5/K0 III       &  14.7  & 0.77 &    0.08 &   6.36&---  & Cool    \\
 13118 & VY Ari       & K3-4 V-IV       &  13.2  & 0.96 &    3.72 &   6.53&---  &         \\
 16042 & UX Ari       & G5 V/K0 IV      &   6.4  & 0.88 &    2.96 &   6.78&---  & Cool    \\
 16846 & V711 Tau     & G5 IV/K1 IV     &   2.8  & 0.88 &    3.51 &   6.84&---  & Both    \\
 19431 & EI Eri       & G5 IV           &   1.9  & 0.71 &    3.28 &   6.73&---  &         \\
 23743 & BM Cam       & K0 III          &  80.9  & 1.11 &   -0.33 &   6.37&---  &         \\
 35600 & AR Mon       & G8 III/K2-3 III &  21.2  & 1.06 &    1.52 &   6.39&---  & Both    \\
 37629 & $\sigma$ Gem & K1 III          &  19.6  & 1.12 &    1.36 &   6.36&---  &         \\
 46159 & IL Hya       & G8 V/K0 III-IV  &  12.9  & 1.01 &    1.96 &   6.53&---  & Cool    \\
 56851 & V829 Cen     & G5 V/K1 IV      &  11.7  & 0.95 &    2.44 &   6.57&---  &         \\
 59600 & HU Vir       & K0 IV           &  10.4  & 0.97 &    3.21 &   6.63&---  &         \\
 59796 & DK Dra       & K1 III/K1 III   &  64.4  & 1.15 &    0.59 &   6.33&---  & Both    \\
 65187 & BM CVn       & K1 III          &  20.6  & 1.16 &    2.08 &   6.41&---  &         \\
 82080 & $\varepsilon$ UMi & A8-F0 V/G5 III& 39.5 &0.90 &   -0.92 &   5.99&---  & Cool    \\
 84586 & V824 Ara     & G5 IV/K0 V-IV   &   1.7  & 0.80 &    4.38 &   6.89&---  & Both    \\
 85852 & DR Dra       & WD/K0-2 III     & 905.9  & 1.04 &    1.54 &   6.31&---  & Cool    \\
 94013 & V1762 Cyg    & K1 IV-III       &  28.6  & 1.09 &    1.65 &   6.40&---  &         \\
 95244 & V4138 Sgr    & K1 III          &  13.0  & 1.03 &    2.00 &   6.39&---  &         \\
 96003 & V1817 Cyg    & A2 V/K2 III-II  & 108.8  & 1.12 &   -1.17 &   6.36&---  & Cool    \\
 96467 & V1764 Cyg    & F-K1 III:       &  40.1  & 1.22 &    0.45 &   6.33&---  &         \\
109002 & HK Lac       & F1 V/K0 III     &  24.4  & 1.05 &    1.02 &   6.57&---  & Cool    \\
109303 & AR Lac       & G2 IV/K0 IV     &   2.0  & 0.76 &    2.99 &   6.51&-0.70& Both    \\
111072 & V350 Lac     & K2 III          &  17.7  & 1.17 &    0.97 &   6.03&---  &         \\
112997 & IM Peg       & K2 III-II       &  24.4  & 1.13 &    0.93 &   6.43&---  &         \\
114639 & SZ Psc       & F8 IV/K1 IV     &   4.0  & 0.79 &    2.67 &   6.71&---  & Cool    \\
116584 & $\lambda$ And & G8 III-IV      &  20.5  & 0.98 &    1.75 &   6.48&-0.56&         \\
117915 & II Peg       & K2-3 V-IV       &   6.7  & 1.01 &    4.38 &   6.68&---  &         \\

\noalign{\smallskip} \hline \noalign{\smallskip}
\end{tabular}
\end{flushleft}
\par
\vspace{0.3cm} \noindent {\it Notes}: {Spectral types and orbital periods are from the
Catalog of Active Binary Stars (CABS) by \cite {strass} except for 
IL Hya (see \cite {weber}).\\
The last Column is according to CABS except for TZ Tri 
and IL Hya (see \cite{montes} and \cite{fekel} respectively).}

\label{tab:data2}
\end{table*}

\begin{table*}
\caption{\ion{Mg}{ii} {\it k} line fluxes for normal stars. 
For each star we give the Hipparcos number,
the spectral type and luminosity class, the $B-V$ color index,
the absolute visual magnitude, the log of the surface \ion{Mg}{ii} {\it k} flux,
and the metallicity.}
\begin{flushleft}
\begin{tabular}{| r | l c c c c r r r| r | l c c c c |}
\hline \hline \noalign{\smallskip}

HIP&$Sp.Type$ &  $B-V$& $M_{\it V}$&  $log~F_{\ion{Mg}{ii}}$& [Fe/H]& & & & 
HIP&$Sp.Type$ &  $B-V$& $M_{\it V}$&  $log~F_{\ion{Mg}{ii}}$& [Fe/H]\\

\noalign{\smallskip}
\hline \noalign{\smallskip}
   544   &K0 V    &  0.75 &   5.39 &    6.10&  ---         &&&&  31592   &K0 III  &  1.04 &   2.46 &    5.20&   .05\\
  1562   &K2 III  &  1.21 &  -1.18 &    4.88&  -.09        &&&&  32768   &K0 III  &  1.21 &  -0.80 &    4.79&  ---\\ 
  2021   &G2 IV   &  0.62 &   3.45 &    5.26&  -.23        &&&&  33817   &K1 V    &  0.88 &   5.89 &    5.93&  ---\\ 
  2081   &K0 III  &  1.08 &   0.52 &    5.14&  ---         &&&&  34622   &K0 III  &  1.02 &   0.86 &    5.18&  ---\\ 
  3092   &K3 III  &  1.27 &   0.81 &    4.71&   .04        &&&&  35264   &K3 I    &  1.62 &  -4.92 &    4.48&  ---\\ 
  3093   &K0 V    &  0.85 &   5.65 &    5.45&  -.17        &&&&  35907   &K0 III  &  1.25 &  -1.51 &    4.96&   .06\\
  3179   &K0 II   &  1.17 &  -1.99 &    4.96&  -.09        &&&&  37379   &K3 III  &  1.54 &  -1.61 &    4.23&  -.22\\
  3419   &K0 III  &  1.02 &  -0.30 &    5.50&   .13        &&&&  37447   &K0 III  &  1.02 &   0.71 &    4.95&  -.09\\
  3765   &K2 V    &  0.89 &   6.38 &    5.50&  -.29        &&&&  37740   &G8 III  &  0.93 &   0.35 &    5.13&  -.16\\
  3821   &G0 V    &  0.59 &   4.59 &    5.60&  -.31        &&&&  37826   &K0 III  &  0.99 &   1.09 &    5.10&  -.04\\
&&&&&&&&&&&&&&\\
  4422   &G8 III  &  0.96 &   0.62 &    5.05&  -.51        &&&&  38170   &G6 I    &  1.22 &  -4.74 &    5.31&   .24\\
  5364   &K2 III  &  1.16 &   0.68 &    4.88&   .04        &&&&  40526   &K4 III  &  1.48 &  -1.22 &    4.38&  -.24\\
  5447   &M0 III  &  1.58 &  -1.86 &    4.63&  -.10        &&&&  41704   &G4 II   &  0.86 &  -0.40 &    5.28&  -.21\\
  6537   &K0 III  &  1.07 &   0.87 &    4.96&  -.22        &&&&  41926   &K0 V    &  0.78 &   5.95 &    5.58&  --- \\
  6867   &K5 II   &  1.54 &  -0.87 &    4.82&  ---         &&&&  42673   &K0 III  &  0.96 &  -0.35 &    5.57&  --- \\
  7607   &K3 III  &  1.27 &  -0.04 &    4.67&   .00        &&&&  42808   &K2 V    &  0.92 &   6.35 &    6.08&  --- \\
  7884   &K3 III  &  1.35 &  -0.81 &    4.48&  -.24        &&&&  43813   &G8 III  &  0.98 &  -0.21 &    5.08&   .38\\
  8102   &G8 V    &  0.73 &   5.68 &    5.48&  -.66        &&&&  44700   &G8 I    &  1.04 &  -2.03 &    5.46&   .38\\
  9884   &K2 III  &  1.15 &   0.48 &    4.95&  -.21        &&&&  44897   &F9 V    &  0.58 &   4.54 &    5.77&  --- \\
 10644   &G0 V    &  0.61 &   4.66 &    6.05&  -.30        &&&&  45860   &M0 III  &  1.55 &  -1.02 &    4.58&  -.26\\
&&&&&&&&&&&&&&\\
 12114   &K3 V    &  0.92 &   6.50 &    5.42&  ---         &&&&  46390   &K3 III  &  1.44 &  -1.69 &    4.61&  -.12\\
 12444   &F6 V    &  0.52 &   4.12 &    5.95&   .02        &&&&  46404   &G2 V    &  0.64 &   2.91 &    5.62&  -.31\\
 13701   &K1 III  &  1.09 &   0.83 &    4.98&  -.23        &&&&  47193   &K3 III  &  1.49 &  -3.31 &    4.64&   .09\\
 14135   &M2 III  &  1.63 &  -1.61 &    4.26&  ---         &&&&  47908   &G0 II   &  0.81 &  -1.46 &    5.59&   .17\\
 14668   &K0 III  &  0.98 &   1.11 &    5.11&   .04        &&&&  48455   &K0 III  &  1.22 &   0.83 &    4.91&   .12\\
 15474   &M3 III  &  1.61 &  -0.79 &    4.68&  ---         &&&&  50583   &K0 III  &  1.13 &  -0.92 &    4.43&  -.49\\
 16134   &K5 V    &  1.34 &   7.89 &    5.74&  ---         &&&&  51172   &K4 III  &  1.43 &  -0.97 &    4.72&  -.39\\
 17440   &K0 IV   &  1.13 &   1.41 &    4.92&  ---         &&&&  52727   &G5 III  &  0.90 &  -0.06 &    5.81&  --- \\
 17678   &M2 III  &  1.59 &  -0.83 &    4.61&  ---         &&&&  53229   &K0 III  &  1.04 &   1.41 &    4.61&  -.20\\
 18543   &M1 III  &  1.59 &  -1.19 &    4.52&  ---         &&&&  53721   &G0 V    &  0.62 &   4.29 &    5.39&   .01\\
&&&&&&&&&&&&&&\\
 19335   &F7 V    &  0.52 &   3.87 &    6.24&  ---         &&&&  53740   &K1 III  &  1.08 &   0.44 &    4.98&  -.22\\
 19747   &K1 III  &  1.09 &   1.07 &    4.99&  ---         &&&&  54539   &K1 III  &  1.14 &  -0.27 &    4.91&  -.13\\
 19921   &K2 IV   &  1.08 &   3.13 &    5.01&  ---         &&&&  55282   &K0 III  &  1.11 &  -0.32 &    5.05&  -.48\\
 20205   &G8 III  &  0.98 &   0.28 &    5.44&   .13        &&&&  57439   &G0 II   &  0.89 &  -1.51 &    5.57&  --- \\
 20455   &G8 III  &  0.98 &   0.41 &    5.23&   .06        &&&&  57757   &F8 V    &  0.52 &   3.40 &    5.88&   .13\\
 20850   &K0 V    &  0.84 &   5.66 &    6.08&  ---         &&&&  58576   &K0 IV   &  0.76 &   4.99 &    5.55&   .16\\
 20885   &G7 III  &  0.95 &   0.42 &    5.54&   .04        &&&&  59316   &K2 III  &  1.33 &  -1.82 &    4.20&  -.13\\
 20889   &K0 III  &  1.01 &   0.14 &    5.04&   .04        &&&&  60172   &K1 III  &  1.17 &   0.26 &    5.18&  -.48\\
 20917   &K7 V    &  1.36 &   8.00 &    5.70&  ---         &&&&  60260   &K3 III  &  1.39 &  -0.63 &    4.74&  --- \\
 22263   &G3 V    &  0.63 &   4.87 &    6.19&  -.13        &&&&  61084   &M4 III  &  1.60 &  -0.56 &    4.62&  --- \\
&&&&&&&&&&&&&&\\
 22449   &F6 V    &  0.48 &   3.67 &    6.05&   .02        &&&&  61359   &G5 II   &  0.89 &  -0.51 &    5.30&   .27\\
 23123   &K2 II   &  1.37 &  -2.86 &    4.67&   .26        &&&&  63090   &M3 III  &  1.57 &  -0.57 &    4.72&  -.09\\
 26366   &G8 III  &  0.95 &   1.33 &    5.38&  -.53        &&&&  64022   &K5 III  &  1.48 &  -0.04 &    4.43&  -.26\\
 27628   &K1 III  &  1.15 &   1.01 &    4.89&   .13        &&&&  64394   &G0 V    &  0.57 &   4.42 &    5.91&   .03\\
 27750   &K2 II   &  1.38 &  -2.91 &    4.87&  -.25        &&&&  64792   &G0 V    &  0.58 &   3.92 &    6.35&   .10\\
 27890   &K1 III  &  1.02 &   2.47 &    4.90&   .10        &&&&  64962   &G8 III  &  0.92 &  -0.05 &    5.20&   .06\\
 27913   &G0 V    &  0.59 &   4.70 &    6.47&  -.03        &&&&  65721   &G5 V    &  0.71 &   3.68 &    5.25&  -.11\\
 29696   &G8 III  &  1.02 &   0.75 &    5.15&  -.33        &&&&  67275   &F7 V    &  0.51 &   3.54 &    6.06&   .00\\
 30343   &M3 III  &  1.62 &  -1.39 &    4.43&  ---         &&&&  67457   &M5 III  &  1.52 &   0.51 &    4.96&  --- \\
 31205   &K0 III  &  1.10 &   0.88 &    5.34&  -.42        &&&&  68815   &M6 III  &  1.24 &   0.68 &    4.47&  --- \\

\noalign{\smallskip} \hline \noalign{\smallskip}
\end{tabular}
\end{flushleft}
\par
\label{tab:data3}
\end{table*}

\begin{table*}
{\bf Table 3.} (continued)
\begin{flushleft}
\begin{tabular}{| r | l c c c c r r r| r | l c c c c |}
\hline \hline \noalign{\smallskip}

HIP&$Sp.Type$ &  $B-V$& $M_{\it V}$&  $log~F_{\ion{Mg}{ii}}$& [Fe/H]& & & &
HIP&$Sp.Type$ &  $B-V$& $M_{\it V}$&  $log~F_{\ion{Mg}{ii}}$& [Fe/H]\\
\noalign{\smallskip}
\hline \noalign{\smallskip}

 68933   &K0 III  &  1.01 &   0.70 &    5.05&   .03        &&&&  90496   &K1 III  &  1.02 &   0.95 &    5.11&  -.20\\
 70692   &K4 III  &  1.43 &  -0.87 &    4.72&  -.16        &&&&  91117   &K2 III  &  1.32 &   0.21 &    4.62&  -.18\\
 71053   &K4 III  &  1.30 &   0.27 &    4.88&  -.17        &&&&  92043   &F6 V    &  0.48 &   2.79 &    5.97&  -.11\\
 71681   &K1 V    &  0.90 &   5.70 &    5.59&   .26        &&&&  92761   &K1 II   &  1.41 &  -3.91 &    5.06&  --- \\
 71683   &G2 V    &  0.71 &   4.34 &    5.37&   .15        &&&&  92791   &M4 III  &  1.58 &  -2.98 &    4.79&  --- \\
 72010   &K3 III  &  1.36 &   0.07 &    4.70&  ---         &&&&  92862   &M5 III  &  1.40 &  -1.07 &    5.25&  --- \\
 72370   &K5 III  &  1.43 &  -1.67 &    4.46&  ---         &&&&  93864   &K1 III  &  1.17 &   0.48 &    5.02&  -.23\\
 73184   &K4 V    &  1.02 &   6.86 &    5.64&   .01        &&&&  94376   &G9 III  &  0.99 &   0.63 &    5.09&  -.27\\
 74395   &G8 III  &  0.92 &   0.65 &    5.09&  ---         &&&&  94713   &K0 II   &  1.26 &  -2.51 &    4.92&   .28\\
 74666   &G8 III  &  0.96 &   0.69 &    5.23&  -.26        &&&&  95822   &K0 III  &  1.05 &   0.85 &    4.89&  -.08\\
&&&&&&&&&&&&&&\\
 76423   &M5 II   &  1.20 &   0.50 &    5.56&  ---         &&&&  97433   &G8 III  &  0.89 &   0.59 &    5.23&  -.18\\
 77070   &K2 III  &  1.17 &   0.87 &    4.70&  -.05        &&&&  98110   &K0 II   &  1.02 &   0.74 &    5.09&  -.09\\
 79882   &G8 III  &  0.97 &   0.64 &    5.05&  -.01        &&&&  99461   &K2 V    &  0.87 &   6.41 &    5.50&  -.58\\
 80331   &G8 III  &  0.91 &   0.58 &    5.20&  -.21        &&&& 100437   &K0 II   &  1.08 &   0.06 &    5.25&   .00\\
 80704   &M6 III  &  1.29 &  -0.39 &    5.14&   .02        &&&& 101474   &K2 I    &  1.59 &  -2.66 &    4.60&  --- \\
 80816   &G8 III  &  0.95 &  -0.50 &    4.41&  -.27        &&&& 101772   &K0 III  &  1.00 &   0.65 &    5.03&   .03\\
 81065   &G9 IV   &  0.92 &   0.41 &    5.62&  -.05        &&&& 102422   &K0 IV   &  0.91 &   2.63 &    5.27&  -.32\\
 81833   &G8 III  &  0.92 &   0.80 &    5.45&  -.18        &&&& 102488   &K0 III  &  1.02 &   0.76 &    5.06&  -.18\\
 82273   &K2 II   &  1.45 &  -3.62 &    4.89&  -.06        &&&& 103227   &K0 III  &  1.25 &  -2.66 &    5.15&  -.06\\
 82396   &K2 III  &  1.14 &   0.78 &    4.92&  -.17        &&&& 104060   &K5 I    &  1.61 &  -4.07 &    4.54&  -.45\\
&&&&&&&&&&&&&&\\
 84345   &M5 II   &  1.16 &  -2.57 &    5.55&  ---         &&&& 105090   &K7 V    &  1.40 &   8.71 &    5.49&  --- \\
 84380   &K3 II   &  1.44 &  -2.10 &    4.62&  -.18        &&&& 105406   &F8 V    &  0.53 &   4.27 &    6.25&  -.13\\
 85068   &G8 III  &  0.99 &   0.27 &    5.37&  ---         &&&& 106642   &M4 III  &  1.34 &  -0.43 &    4.21&  --- \\
 85258   &K3 I    &  1.48 &  -3.49 &    5.02&   .50        &&&& 107089   &K0 III  &  1.01 &   2.10 &    5.21&  --- \\
 85670   &G2 II   &  0.95 &  -2.43 &    5.98&   .14        &&&& 107119   &K0 III  &  1.11 &   0.89 &    4.90&   .04\\
 86742   &K2 III  &  1.17 &   0.76 &    4.84&   .00        &&&& 107472   &K0 I    &  1.38 &  -2.41 &    5.46&   .06\\
 87158   &G5 IV   &  0.94 &   1.27 &    5.68&  ---         &&&& 108870   &K5 V    &  1.06 &   6.89 &    5.79&  -.23\\
 87261   &K0 III  &  1.19 &   0.24 &    4.87&  ---         &&&& 109492   &K1 I    &  1.56 &  -3.35 &    4.81&   .22\\
 87585   &K2 III  &  1.18 &   1.06 &    4.92&  -.09        &&&& 109937   &K3 III  &  1.45 &  -2.28 &    4.48&  -.12\\
 87808   &K1 II   &  1.35 &  -2.70 &    4.96&  -.24        &&&& 110130   &K3 III  &  1.39 &  -1.05 &    4.63&  --- \\
&&&&&&&&&&&&&&\\
 87933   &G8 III  &  0.94 &   0.61 &    5.62&  -.10        &&&& 112724   &K0 III  &  1.05 &   0.76 &    5.03&   .02\\
 88048   &G9 III  &  0.99 &  -0.03 &    4.93&   .16        &&&& 112748   &G8 III  &  0.93 &   0.74 &    5.07&  -.03\\
 88601   &K0 V    &  0.86 &   5.50 &    5.96&  -.05        &&&& 112961   &M2 III  &  1.63 &  -1.67 &    4.53&  --- \\
 88839   &G7 III  &  0.94 &  -0.59 &    5.13&  ---         &&&& 113881   &M2 II   &  1.65 &  -1.49 &    4.25&  -.11\\
 88972   &K2 V    &  0.88 &   6.15 &    5.54&  -.20        &&&& 114971   &G7 III  &  0.92 &   0.68 &    5.20&  -.44\\
 89962   &K0 III  &  0.94 &   1.84 &    5.18&  -.42        &&&& 116727   &K1 IV   &  1.03 &   2.51 &    5.20&   .04\\
 90139   &K2 III  &  1.17 &   0.87 &    4.93&  -.16        &&&&          &        &       &        &        &      \\
\noalign{\smallskip} \hline \noalign{\smallskip}
\end{tabular}
\end{flushleft}
\par
\vspace{0.3cm} \noindent { Units for the \ion{Mg}{ii} surface fluxes $F$ are in
$erg~cm^{-2}~s^{-1}$.  }
\label{tab:data4}
\end{table*}

The spectroscopic \ion{Mg}{ii} observations have been obtained
with the {\it IUE} high resolution long wavelength camera.
The calibrated spectra have been analysed and processed in order
to obtain the \ion{Mg}{ii} {\it k} emission line strength
using the method described in Paper II, which consists of a direct
integration of the observed profiles, as measured above the 
photospheric flux level.
As stars in our sample all have parallaxes measured astrometrically, they
all are nearby stars. This makes us confident that the interstellar
absorbtion can be neglected. On the other hand the evaluation of the
interstellar absorbtion requires information which is usually lacking.
The estimated relative uncertainty in the derivations of the apparent 
flux from observations is ${\pm} 15\%$.
In the case of binaries with two
emitting stars (three known objects between BY Dra and five between RS CVn)
we have not deblended both contributions.
Indeed AR Lac is the only target in our sample in which the individual
contribution from the two stars is evident as
a consequence of their similar \ion{Mg}{ii} luminosity and of
the favorable inclination angle of the system.
Due to the statistical character of the present investigation,
we have not considered it necessary to perform a detailed analysis of
these spectra so that the integrated flux of both
componets is considered.

In our data sample there are stars that have been observed two or more times.
Some active stars have been observed hundreds of times.
In all these cases a mean value of the flux was computed.
 
The apparent visual magnitudes of all stars along with their 
luminosity class, $B-V$, trigonometric parallaxes and the relative errors
were obtained mostly from the {\it HIPPARCOS} catalogue. In a few cases,
mainly regarding RS CVn and BY Dra stars,
information on colour and luminosity class was taken from the 
Strasbourgh Data Center (SIMBAD) or from the recent literature.
We have assumed a typical error of ${\pm} 0.01$ mag on visual magnitudes.

To convert fluxes at Earth, $f$, into stellar surface fluxes, $F$, we have used the
following relation:

\begin{equation}
\rm {log~(F/f) = 0.350 + 4~log~T_{eff} + 0.4 (V + BC)}
\label{eq:flux}
\end{equation}

\noindent where $T_{eff}$ is the effective temperature and BC the bolometric
correction both derived from $B-V$ colour using the coefficients to the
polinomial fits given by \cite{flow}; $V$ is the apparent visual magnitude.
The constant 0.350 is derived from the solar values given by \cite{allen}.
The uncertainty on the chromospheric surface fluxes resulting from
this conversion is hardly evaluable and the cooler the stars
the larger the uncertainty.
For binaries, we used combined values of $B-V$; in this way
a further error in the calculation of surface fluxes is introduced.
To prevent too large uncertainties in BC we have limited our sample to 
$B-V<1.7$ which corresponds to $BC\approx-2.3$.

Information on metallicity ([Fe/H]) could be found only
for a subsample of 130 quiet stars and few very active stars 
(see \cite{carney}; \cite{cayrel}). 

Table \ref{tab:data1} and table \ref{tab:data2} give the \ion{Mg}{ii} {\it k}
logarithmic surface fluxes, together with other relevant data, for BY Dra and
RS CVn stars respectively.                                   
Similar information is provided in Table \ref{tab:data3} for normal stars.
Units for the \ion{Mg} {ii} surface fluxes $F$ are in
$erg~cm^{-2}~s^{-1}$.

\section{Discussion of the observations}
\label{sec:discu}

In this Section we investigate the dependence, if any, of the
chromospheric radiative loss rates on fundamental stellar parameters
such as luminosity, metallicity, and temperature. 

\subsection{Line intensity--luminosity relation}
\label{sec:lumin}

\begin{figure}
\includegraphics[width=8.8cm]{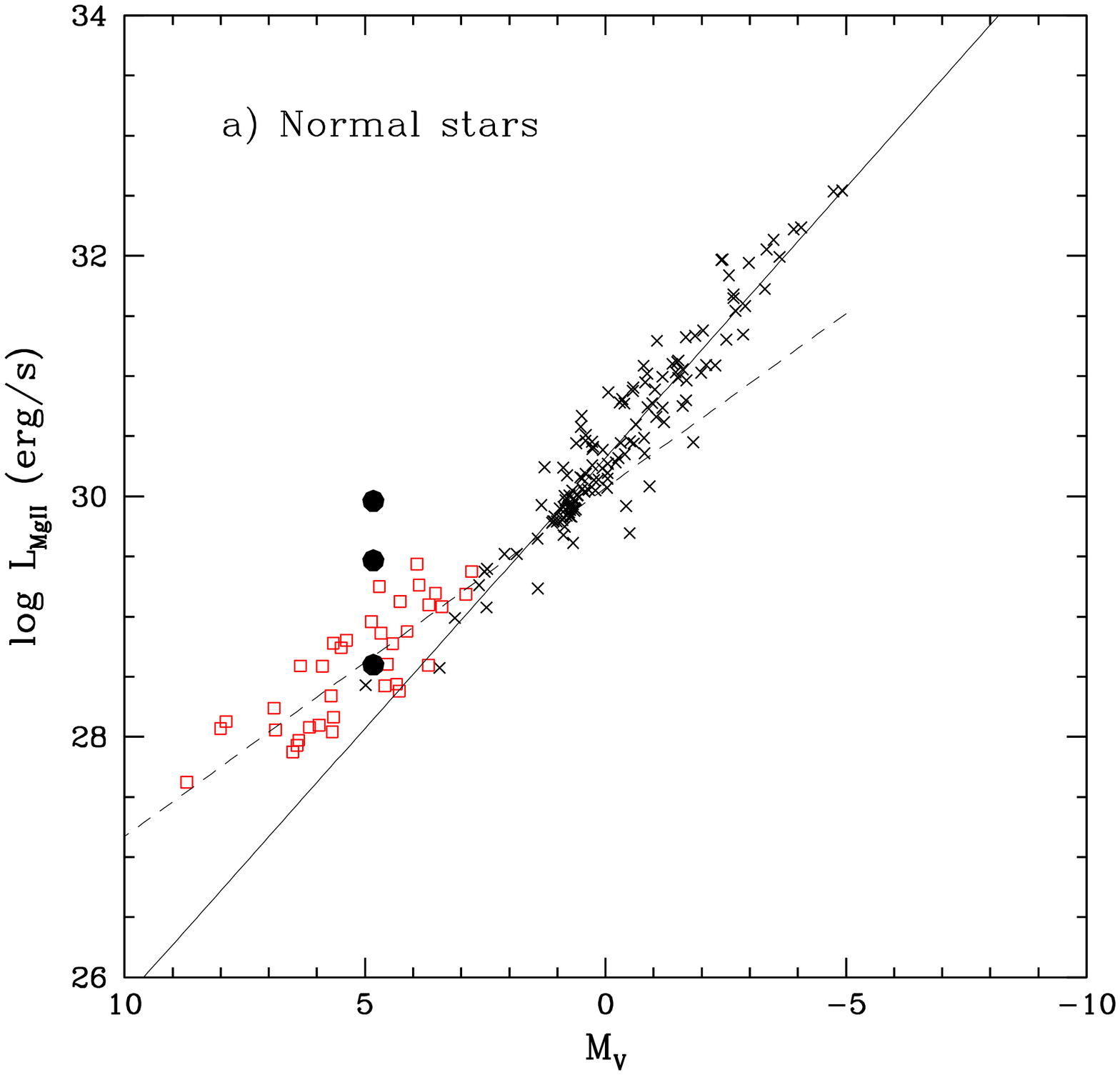}
\includegraphics[width=8.8cm]{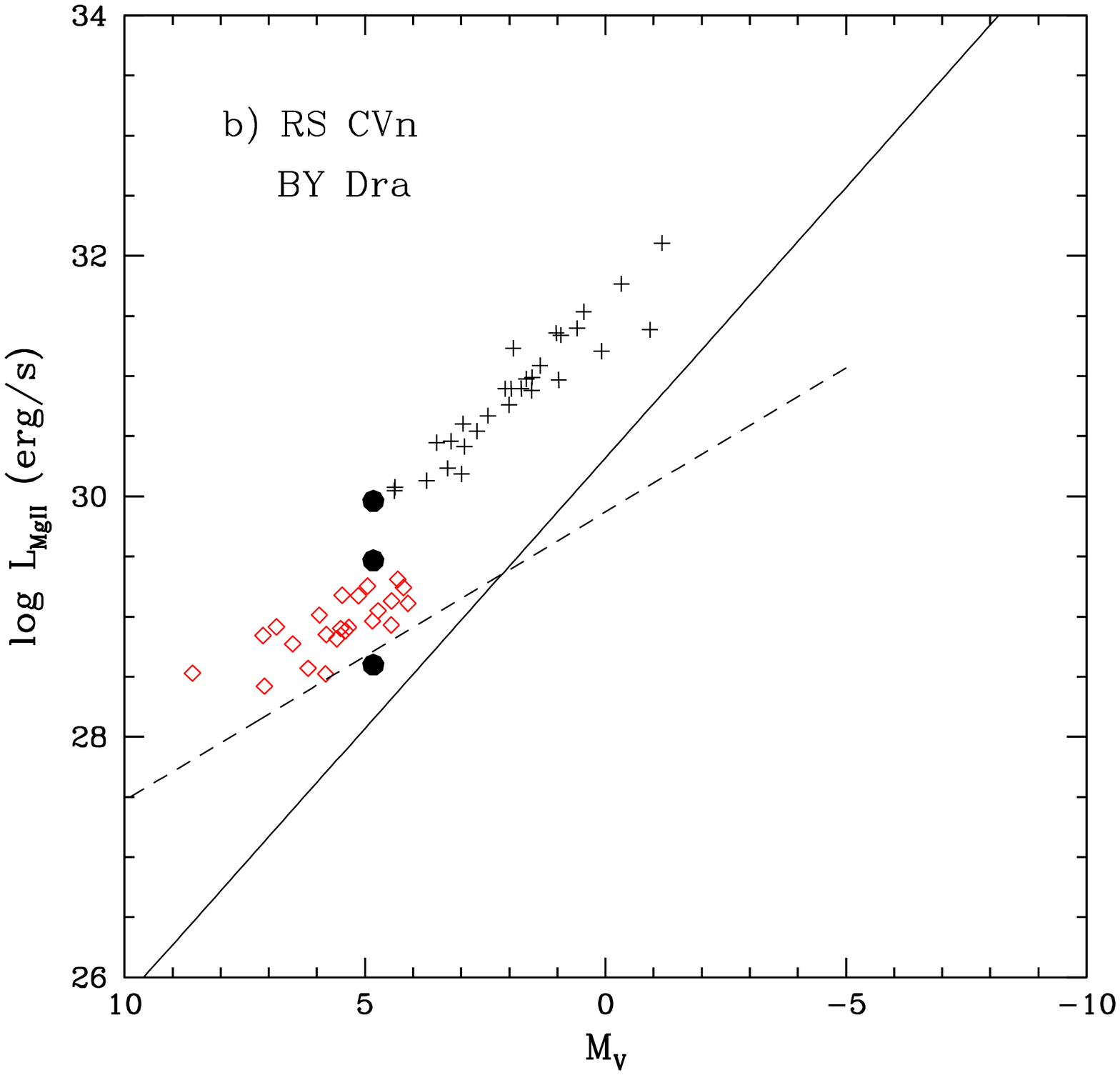}
\caption{ The logarithmic absolute luminosity in the \ion{Mg}{ii} {\it k}
line is plotted as a function of the absolute visual magnitude.
(a) Quiet stars. Dwarfs are shown as open squares and evolved stars as crosses.
(b) RS CVn (pluses) and BY Dra stars (diamonds). 
The best fits to the data of normal stars are represented 
for dwarfs by a dashed line and for giants by a full line.
For comparison also data of the Sun are reported as filled circles (see text).
}
\label{fig:magint}
\end{figure}

In Fig. \ref {fig:magint} we have plotted the logarithm 
of the \ion{Mg}{ii} {\it k} line absolute luminosity 
against absolute magnitude for normal stars (Fig. \ref {fig:magint}a) and
active stars (Fig. \ref {fig:magint}b). 
Also indicated are the linear fits
to the data for, separately, evolved (crosses) and main sequence 
(open squares) normal stars.
To simplify the figures, 
error bars are not reported. 
In the figure are also drawn, for comparison, the "average" 
\ion{Mg}{ii} {\it k} luminosity
for quiet solar regions, the "average" \ion{Mg}{ii} {\it k} luminosity for
plage areas and the \ion{Mg}{ii} {\it k} luminosity in the flare of September
5, 1973, as reported by Cerruti--Sola et al. (1992) (filled circles).

Fig. \ref {fig:magint}a clearly shows that the dependence of the 
line intensity on the
stellar luminosity is different for evolved and dwarf normal stars. 
A least--squares fit to the data of normal evolved stars (class I--IV)
provides:

\begin{equation}
\rm {log~L_{\ion{Mg}{ii}} = (30.32{\pm} 0.01) - (0.45{\pm} 0.01)~ M_{\it V}}
\label{eq:lum}
\end{equation}

\noindent with a correlation coefficient of 0.95. 
The slope coefficient of about -0.4 in equation (\ref{eq:lum}) suggests that 
for giants and supergiants the \ion{Mg}{ii} line intensity is directly
proportional to stellar luminosity.

A similar fitting procedure applied to normal dwarf stars gives: 

\begin{equation}
\rm {log~L_{\ion{Mg}{ii}} = (30.07{\pm} 0.04) - (0.29{\pm} 0.01)~ M_{\it V}}
\label{eq:lumd}
\end{equation}

\noindent with a correlation coefficient of 0.82.
The slope coefficient smaller than in Eq. (\ref{eq:lum}) 
indicates that dwarf stars 
have an excess of \ion{Mg}{ii} {\it k}
luminosity which increases with increasing magnitude.

Eq. (\ref{eq:lum}) is in very good agreement with the 
result of Weiler \& Oegerle (1979)
based on stars brighter than $M_{\it V}$ = 4 but it
is quite different from our fit in Paper II because this last was performed
on the whole sample of dwarfs and giants together. 
Other authors
(e.g. \cite{linskyayres}; \cite{weiler}; \cite{basri})
report, on the basis of a sample mainly containing
giant and supergiant stars and only few dwarfs, that the ratio of 
the \ion{Mg}{ii} flux rate to stellar 
luminosity is not dependent on stellar luminosity.
On the basis of the present analysis, we confirm that this is true only 
for evolved stars
but is not true for main sequence stars.

As far as very active stars are concerned, we show in Fig. \ref {fig:magint}b
the location of RS CVn + BY Dra in the ($log~L_{\ion{Mg}{ii}}$, $M_{\it V}$)
diagram.
We can notice that 
there is a good correlation between the logarithmic 
\ion{Mg}{ii} {\it k} line intensity and $M_{\it V}$ 
for RS CVn stars (pluses).
A fit to these data gives:

\begin{equation}
\rm {log~L_{\ion{Mg}{ii}} = (31.46{\pm} 0.04) - (0.33{\pm} 0.01)~ M_{\it V}}
\label{eq:lumrs}
\end{equation}

\noindent with r=0.94.

Fig. \ref {fig:magint}b also shows that BY Dra stars 
(diamonds) have a slope similar to that of 
normal main sequence stars.
However, the coefficients of the fit are not reported here
given the quite low data correlation (r=0.74).

The mean activity level of RS CVn 
stars is comparable to that of an intense solar flare (about 1.3 dex greater
than the level of quiet Sun) and that the activity level of BY Dra stars
may be as large as that of solar plages (the mean run exceeds
by 0.6 dex the quiet Sun).

\subsection{ Flux--metallicity relation}
\label{sec:metal}

\begin{figure}
\includegraphics[width=8.8cm]{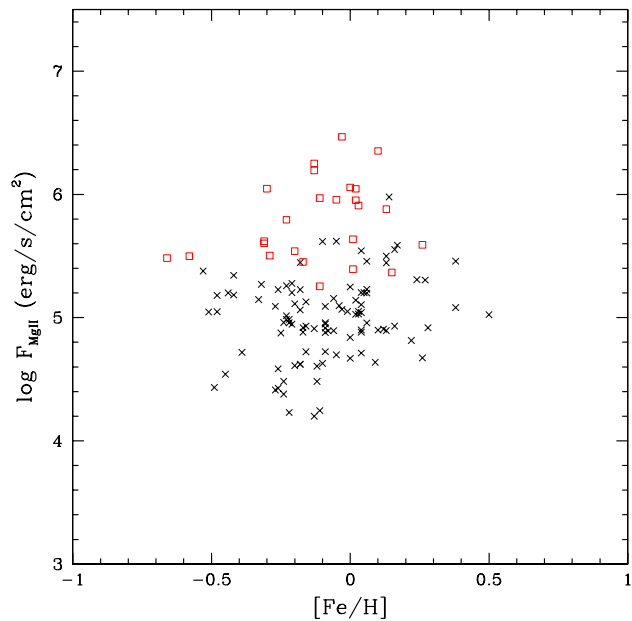}
\caption{ The logarithm of the surface flux in the \ion{Mg}{ii} {\it k}
line is plotted as a function of metallicity for a subsample of 130
stars.  Open squares indicate main sequence stars and crosses indicate evolved stars.
}
\label{fig:metf}
\end{figure}

In Fig. \ref {fig:metf} is plotted the logarithm of the \ion{Mg}{ii} {\it k}
surface flux as a function of metallicity,
for the 130 stars for which this latter parameter is known.
The figure shows clearly that, 
in spite of the large range of metallicity covered,
the radiative losses from the \ion{Mg}{ii} {\it k} line
do not show any correlation with chemical abundance.
This is true also considering evolved (crosses) and main sequence
stars (open squares) separately.
The figure shows also that main sequence stars have greater 
surface fluxes than evolved stars with the same metallicity, thus confirming
the higher level of activity of these stars as already pointed out
in Sec. \ref{sec:lumin}.

Due to the scarcity of the metallicity data available (see
table \ref{tab:data1} and table \ref{tab:data2}), a similar
analysis has not been attempted for BY Dra and RS CVn stars.

\subsection{Flux--colour relation}
\label{sec:color}

\begin{figure*}
\centerline{\hbox{
\includegraphics[width=8.8cm]{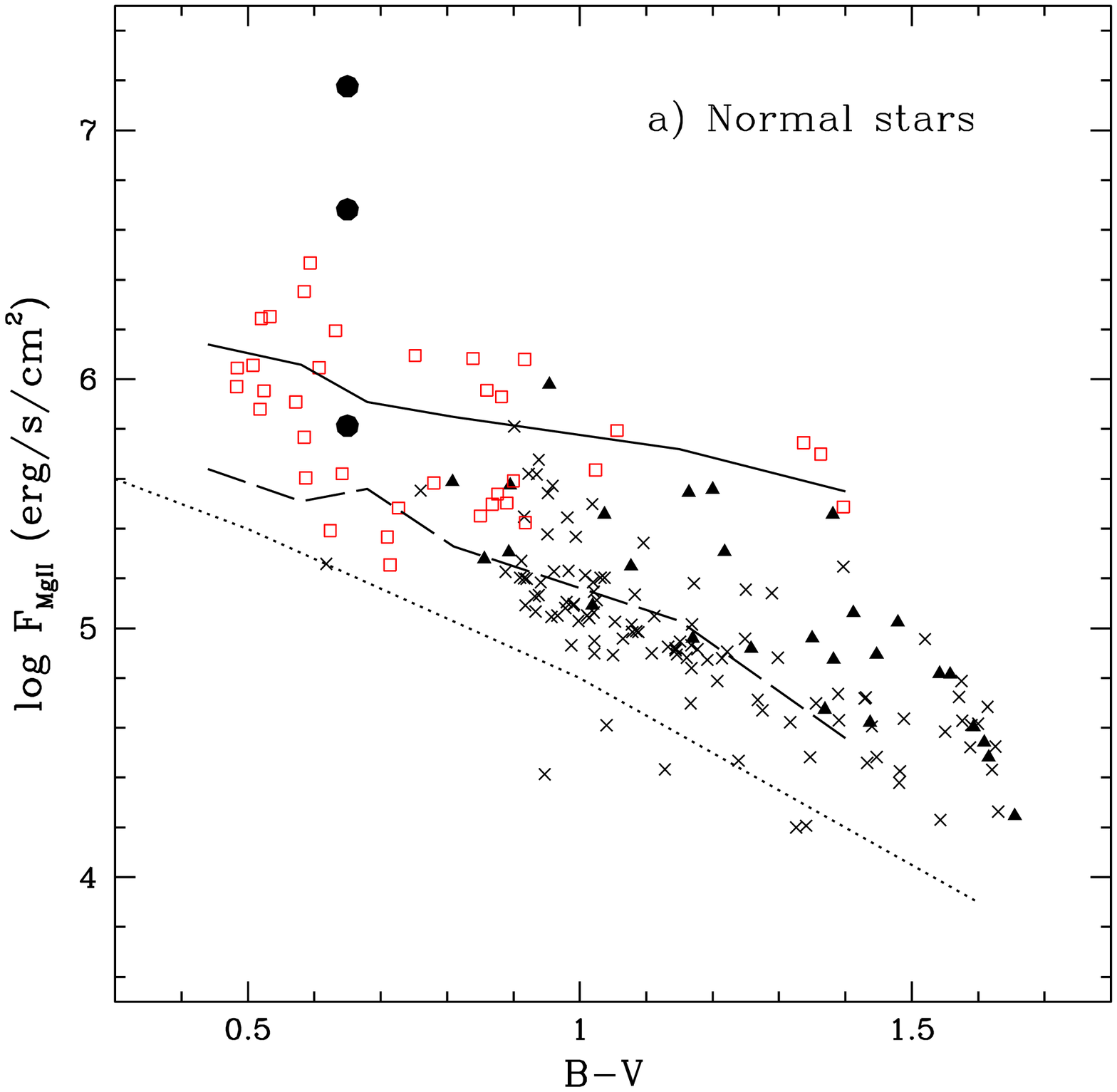}
\includegraphics[width=8.8cm]{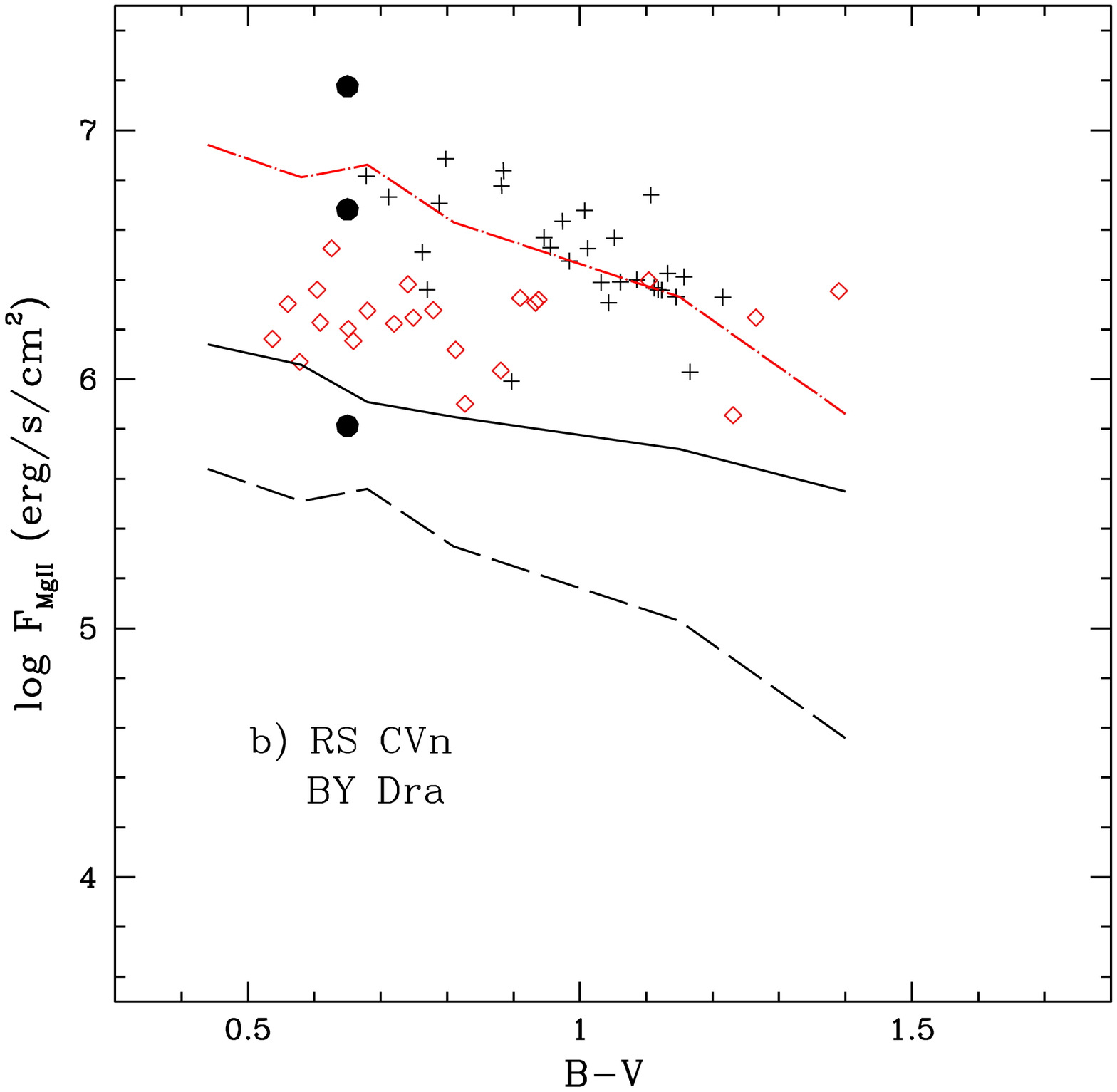}}}
\caption{The logarithm of the observed surface flux in the \ion{Mg}{ii} {\it k}
line is plotted as a function of $B-V$.
(a) Quiet stars. Dwarfs are shown as open squares, class III--IV stars as 
crosses and class I--II stars as triangles.
(b) BY Dra stars (diamonds) and RS CVn (pluses). 
Also shown in each panel are the theoretical fluxes for pure 
acustic wave heating (dashed) 
and for magnetic wave heating in flux tubes (solid) as given by Fawzy et al.
(2002b; Fig. 11) and corrected to consider the {\it k} line only.
The dotted line in panel (a) represent the observational 
minimum limits by Rutten et al. (1991).
The dash--dotted line in panel (b) is the theoretical 
flux for pure acustic wave heating increased by 1.3 dex. 
}
\label{fig:bmv}
\end{figure*}

In Fig. \ref {fig:bmv} we have plotted the observed surface fluxes 
in the \ion{Mg}{ii} {\it k}
line versus $B-V$ for normal stars (Fig. \ref {fig:bmv}a) and
active stars (Fig. \ref {fig:bmv}b).

In Fig. \ref {fig:bmv}a we notice that:
\begin {itemize}

\item The observed minimum fluxes of giant (crosses) and supergiant (triangles)
stars decrease with decreasing temperature.

\item The observed minimum flux level of
supergiants (class I--II) is a factor of two larger in 
$F_{\ion{Mg} {ii}}$ compared to giant stars of the same colour.

\item Surface fluxes  of luminosity class V stars (open squares) are 
only weakly dependent on temperature, if not all.
It is difficult for these stars to identify a line of minimum fluxes 
as they are sparsely distributed
with a spread of about one order of magnitude at high temperatures. 
 
\end{itemize}

As for stars with known very active chromospheres shown in 
Fig. \ref {fig:bmv}b we observe that:
\begin {itemize}
\item RS CVn binaries (pluses) lie far above typical stars
as seen before (Sec. \ref{sec:lumin}).
Minimum fluxes for these stars are well represented by the line of
minimum fluxes defined by giants enhanced by a factor as large as 20.

\item BY Dra stars (diamonds) lie 
0.6 dex above normal main squence stars.
Their observed fluxes do not show any dependence on temperature.
\end{itemize}

\section{Interpretation of the results}
\label{sec:interp}

In Fig. \ref {fig:bmv} we compare our observational determination of
\ion{Mg}{ii} {\it k}
line fluxes at stellar surfaces with the simulated estimates of the same
quantities computed by
Fawzy et al. (2002b) using two--component models of stellar
chromospheres for late--type main sequence stars. 
The dashed line refers to fluxes computed taking into account
pure acustic wave heating only ({\it basal} fluxes).
In addition to this mechanism the models
predict an increase in \ion{Mg}{ii} emission due to magnetic tube waves 
proportional to the fraction of the stellar surface covered by magnetic 
fields (filling factor) (\cite {fawzya}). 
The solid curve in Fig. \ref {fig:bmv} refers to a magnetic filling 
factor of 0.4 which corresponds to 
maximum efficiency in 
magnetic wave heating
and then determines the upper boundary of the chromospheric activity.
Fig. \ref {fig:bmv}a also shows the observational minimum limits 
from Rutten et al. (1991) (dotted line). 
All the quoted curves have been decreased by 
a factor of two to consider the {\it k} line only.

Buchholz et al. (1998) have
simulated chromospheric \ion{Mg}{ii} minimum emission fluxes
both for main sequence stars and giants concluding that
they are essentially the same.
Thus, minimum fluxes computed by Fawzy et al. (2002b) for dwarf stars
can be considered valid also for giants.
The existence of such a common lower boundary in the surface flux versus 
colour is strongly supported by the observations of the 
\ion{Ca}{ii} and \ion{Mg}{ii} lines 
(\cite{schrijver}; \cite{mathiou}; \cite{rutten}; \cite {fawzyb}). 

The \ion{Ca}{ii} theoretical low and upper emission 
boundaries agree with the observations
within a factor of two, which is justified given the inaccuracies  
in the model and/or by uncertainties in observed flux determinations.
For \ion{Mg}{ii}, Fawzy et al. (2002b),
comparing their theoretical emission fluxes with observational
fluxes given by Rutten et al. (1991), 
concluded that the modeled minimum flux limit was within
a factor of two compatible with observations, but, 
even taking a factor of two uncertainty into account,
the upper limit obtained was persistently lower than the maximum observed
emission.

Looking at our data in Fig. \ref {fig:bmv}a we see that our observational
minimum limit is well defined in the range  
of $B-V$ between 0.85 and 1.5, where numerous giant stars (class III)
are present. This limit is in good agreement with 
the line of acoustic wave heating of Fawzy et al. (2002b) 
(dashed line in Fig. \ref {fig:bmv}), 
both in absolute magnitude (within a factor 
of about 1.5) and in terms of temperature dependence.
At higher temperatures, where
only main sequence stars are present, the minimum flux levels 
are less well defined, 
but they are still compatible with the theoretical lower limit.
Note that the present minimum fluxes lie above the limit given by
Rutten et al. (1991). 
Class I--II stars emission fluxes are 
greater than that of class III stars. 
This fact, not caused by magnetic heating, 
is probably due to their lower surface gravity. 

In our sample of "normal" stars, the maximum values of 
\ion{Mg}{ii} fluxes are observed in main sequence stars.
If, as said before, dwarf and giant stars have the same 
lower boundary, then
the spread of \ion{Mg}{ii} fluxes in the vertical direction is larger
for dwarfs than for giants stars (not considering supergiants).
This means that
dwarfs may have a range of activity levels
greater than giants of the same spectral type.
We can see that 60\% of normal main sequence stars
lie below the theoretical magnetic wave heating line and all but one lie below 
the theoretical upper limit if a factor of two uncertainty
is taken into account.
Evidently a large fraction of the atmosphere of dwarfs 
may be dominated by magnetic heating but their emission can be accounted
for by present models. 

On the contrary the cited two--component models seem to be inadequate to
explain the observed chromospheric radiative loss rates of BY Dra and
RS CVn stars. 

Fawzy et al. (2002a) models describe correctly the chromospheres of
solar--type stars with low and moderate levels of activity.
In addition their models are based on uniformly distributed
magnetic flux tubes.
BY Dra stars are generally attributed deep convection zones
and fast rotation rates that enhance magnetic activity 
and it is known that BY Dra stars have large spotted regions (\cite{stewart};
\cite{baliunas}; \cite{saar}).

In our data shown in Fig. \ref {fig:bmv}b, 
the BY Dra stars have \ion{Mg}{ii} fluxes which show no dependence 
on temperature, and they all lie above the line of magnetic 
wave heating (full line)
with a mean value of $6.3~erg~cm^{-2}~s^{-1}$ (see also Sec. \ref {sec:lumin}).
Looking at the data relative to the Sun reported in Fig. \ref {fig:bmv},
the value of 6.3 corresponds, following
an empirical computation, to a fraction of stellar surface covered 
by plages of about 32\%, if the brightness
of typical solar plages is assumed 
(it is not excluded however that the intrinsic brightness of the plage
regions exceeds that of typical solar plages) (\cite{cerruti}).
Perhaps stronger magnetic fields and a modeling different from
magnetic flux tubes are necessary to account for the peculiarities
of this kind of star.
 
Fig. \ref {fig:bmv}b also shows that RS CVn stars lie far 
above normal-chromosphere stars 
(see also Fig.\ref {fig:magint}b). 
They seem to have a dependence on the effective temperature which follows
the slope of the {\it basal} flux enhanced by a factor as large as 20 
(dash--dotted line in Fig.\ref {fig:bmv}b).
However this feature may be simulated by different orbital
periods.
These chromospherically active stars are members of binary systems
and interactions between the two components could produce excess
heating by completely non-magnetic processes such as mass transfer
and accretion, tidal interactions and resonances. 
These other physical processes could affect chromospheric
activity in addition to magnetic field strength (see for example \cite{glebocki}).

\section{Conclusion}
\label{sec:concl}

In this paper we have
addressed the still unsolved problem of identifying the possible physical
processes responsible for chromospheric heating.
To this purpose we have measured intensities of 
the \ion{Mg}{ii} {\it k} line
in a large number of {\it IUE} spectra and we have carried out 
an analytical study for different types of stars. 
The results obtained can be summarized as follows: 
\begin {itemize}
\item Metallicity does not affect the 
emission flux level of the \ion{Mg}{ii} {\it k} line.
\item The logarithm of the total {\it k} line emission luminosity is 
linearly related to the absolute visual magnitude.
For evolved stars, both normal and RS CVn, 
the correlation is very strong and allows the linear relationships  
in Eq. (\ref{eq:lum}) and (\ref{eq:lumrs}) to be defined.
Main sequence stars, both normal and BY Dra, are less correlated
in the ($M_{\it V}$, $log~L_{\ion{Mg}{ii}}$) plane and show a smaller
slope.
\item There are indications that the \ion{Mg}{ii} {\it k} flux 
increases slowly with decreasing stellar gravity (supergiants).
\item The observed range of chromospheric activity in the \ion{Mg}{ii} {\it k} line 
in normal stars is fully accounted for by current models based
on acoustic and magnetic wave heating in the form of magnetic flux tubes.
\item Peculiar objects, like BY Dra and RS CVn, for which
current models are inadequate to explain the very large strength
of the \ion{Mg}{ii} {\it k} line, require that additional 
heating processes be taken into account.
\end {itemize}

\begin{acknowledgements}
The author is grateful to Dr. A. Cassatella for providing
measurements of \ion{Mg}{ii} {\it k} line intensities and 
for critically reading the manuscript.
I thank the referee for useful comments and suggestions. 
\end{acknowledgements}

\end{document}